# Evidence for Multiferroic Characteristics in $NdCrTiO_5$

J. Saha, G. Sharma and S. Patnaik*

School of Physical Sciences, Jawaharlal Nehru University, New Delhi-110067, India.

e-mail: spatnaik@mail.jnu.ac.in

**Abstract:**

We report $NdCrTiO_5$ to be an unusual multiferroic material with large magnetic field dependent electric polarization.  While magneto-electric coupling in this two magnetic sub-lattice oxide is well established, the purpose of this study is to look for spontaneous symmetry breaking at the magnetic transition. The conclusions are based on extensive magnetization, dielectric and polarization measurements around its antiferromagnetic ordering temperature of 18K. Room temperature X-ray diffraction pattern of $NdCrTiO_5$ reveals that the sample is single phase with an orthorhombic crystal structure that allows linear magneto-electric coupling. DC magnetization measurement shows magnetization downturn at 11K together with a small kink corresponding to the $Co^{+3}$ sub-lattice ordering at ~18K. An anomaly in dielectric constant is observed around the magnetic ordering temperature that increases substantially with increasing magnetic field. Through detailed pyroelectric current measurements at zero magnetic field, particularly as a function of thermal cycling, we establish that $NdCrTiO_5$ is a genuine multiferroic material that is possibly driven by collinear magneto-striction.

**Introduction:**

Multiferroic materials exhibit co-existence of multiply ordered ferroic states [1-4]. The consequent co-dependence of such orderings has been envisaged towards realization of novel memory devices and in spintronics [5]. In the recent past, a set of novel materials that has attracted considerable attention relates to magnetic multiferroics where the ferroelctricity is driven by magnetic structure (e.g. $TbMnO_3$) rather than ab-initio non-centro-symmetric crystal structure (e.g. $BaTiO_3$). The ferroelectric phase in this case is enabled by spontaneous spatial inversion symmetry breaking at the magnetic ordering temperature. In such cases, dielectric constant is generally suppressed with external magnetic field. In this regard a controversial example is $NdCrTiO_5$. While the magnetic structure and the possibility of large magneto-electric (ME) coupling in $NdCrTiO_5$ were studied in early 1970's, the open question that remains is whether this compound is a genuine multiferroic or simply a linear magneto-electric (ME) material. In both cases, the strength of the polarization depends on the strength and orientation of the magnetic field. Experimentally, the distinction between the two can be accessed; a multiferroic would show induction of ferroelectricity even in the absence of external magnetic field where as electric polarization due to magneto-electric coupling could be achieved only in the presence of magnetic field. This general requirement is valid for many of the recently claimed multiferroic systems where the low temperature saturation polarization is typically two orders of magnitude lower than other established multiferroics such as $BiFeO_3$.[6]

What is confirmed in $NdCrTiO_5$ is the emergence of electric polarization around its anti-ferromagnetic (AFM) ordering temperature $T_N \sim 18$ K. However, unlike other spin frustrated multiferroic systems, such as $YMnO_3$ [7], $Ni_3V_2O_8$ [8], or $Y_2CoMnO_6$ [9], the electric polarization and anomaly in dielectric constant in $NdCrTiO_5$ reportedly appears only in the presence of external magnetic field. This feature is reminiscent of linear magneto-electric effect as evidenced in $MnTiO_3$ [11]. Two major factors contribute to large ME effect. In composite materials the ME effect is generated as a product property of magnetostriction and piezoelectricity. In such materials the ME effect depends on the strength of coupling between the magnetic and electric ordering. On the other hand in intrinsic multiferroic materials where the electric (magnetic) field is induced by magnetic (electric) ordering the ME effect largely depends on the corresponding free energy [12]. Two theoretical models appropriate for magnetism driven ferroelectricity have been suggested [10]. In most systems with spiral spin structure e.g $TbMnO_3$ [13, 14], $DyMnO_3$ [15], $Ni_3V_3O_8$ [16-17] etc, the induced ferroelectricity is described by spin current model [18, 19]. The inverse Dzyaloshinkii-Moriya (DM) interaction in such cases drives the onset of ferroelectricity by complex magnetic ordering [20, 21]. Here, the symmetry breaking leads to induction of polarization **P** which can be described by $\boldsymbol{P} \propto \boldsymbol{e_{ij}} \times (\boldsymbol{S_i} \times \boldsymbol{S_j})$, where $\boldsymbol{e_{ij}}$ is unit vector joining the spin moments $\boldsymbol{S_i}$ and $\boldsymbol{S_j}$ [18]. If $NdCrTiO_5$ is a multiferroic, then the spin current model is applicable if both magnetic sublattices of $Cr^{+3}$ and $Nd^{+3}$ ordered separately leading to overall non-collinear spin configuration. On the other hand, a modified magneto-striction driven mechanism, as seen for $Y_2CoMnO_6$, is suggested if only the collinearly ordered (along c-axis) $Cr^{+3}$ sublattice played the dominant role for ME coupling. This would then indirectly conjure the *ab* plane aligned moments of $Nd^+$ through exchange coupling. Such collinear ordering, where DM term vanishes yet the spatial inversion symmetry is broken, is achieved via the magnetosriction phenomenon [22]. In this communication we report a detailed structural, magnetic and dielectric property analysis in $NdCrTiO_5$ towards an understanding of the nature of multiferroicity, ME effect and its correlation with AFM ordering. We establish that $NdCrTiO_5$ is a multiferroic material due to collinear spin ordering.

**Experimental:**

Polycrystalline samples of $NdCrTiO_5$ were prepared by standard solid state reaction method. Stoichiometric amount of $Nd_2O_3$, $Cr_2O_3$, $TiO_2$ were used. The mixture was well ground for several hours in agate mortar-pestle. After grinding, the mixture was sintered for 24 hours in air at 1000°C. After the first sintering, the mixture was again well ground for few hours and then pressed in form of pellets (Dia=10mm, thickness=1mm). These pellets were heated again in air at 1300°C for another 24 hours. Both heating and cooling rates were kept at a rate of 5°C/min. Room temperature X-ray diffraction (XRD) pattern was recorded by using Cu Kα radiation with an X'pert Pro PanAlytical diffractometer. The DC magnetization measurements were done in *Cryogenic* PPMS. The dielectric measurements were done using an Agilent E4980A LCR meter. For dielectric and pyroelectric current measurements, electrodes were prepared on the sample by painting silver paste on the planar surfaces. The pyroelectric measurement was performed by using a Kiethley 6514 electrometer and polarization was derived from the pyroelectric current by integrating over time. The sample was poled each time from well above ordering temperature under various combinations of electric and magnetic field. The poling field was removed at lowest temperature and terminals were sorted for 15 minutes to avoid any role of electrostatic stray charges. The warming ramp rate was fixed at 5K per minute.

**Results and Discussion:**

Figure 1 shows the room temperature XRD pattern of NdCrTiO$_5$ in range of 2θ = 10° to 90°. The refinement done by Fullprof method confirms that the sample has been synthesized in near single phase. It is found that NdCrTiO$_5$ has an orthorhombic structure with space group *Pbam* and the estimated lattice parameters are **a** = 7.5661(4) Å, **b** = 8.6513(5)Å and **c** = 5.8003(3)Å. The goodness of fit indicators are $R_{wp}$ = 5.75, $R_p$ = 7.21 and $\chi^2$ = 1.27. The atomic coordinates, bond length and bond angles are also estimated from the room temperature XRD and are summarized in Table 1. The compound is isomorphous with HoMn$_2$O$_5$. The schematic cell for orthorhombic NdCrTiO$_5$ is shown in Figure 2. The 8 Cr$^{3+}$ atoms (blue) occupy the 4 faces and are stacked along the c axis. Each Cr$^{3+}$ ion is surrounded by oxygen octahedral. Nd$^{3+}$ ions (gray) lie along *ab* plane and the planes formed by Cr$^{3+}$ ions are interspaced between a Nd$^{3+}$ and a Ti$^{4+}$ plane. Ti$^{4+}$ ions (cyan) are surrounded by edge sharing oxygen tetragonal pyramids and lie in *ab* plane between the two planes formed by Cr$^{3+}$ ions. Oxygen pyramids around Ti$^{4+}$ share corner with the oxygen octahedra of Cr$^{3+}$ ions. According to the neutron study by G Buisson[23], 95% of Cr$^{3+}$ ions and 5% of Ti$^{4+}$ ions are distributed at 4*f* sites, while 5% of Cr$^{3+}$ ions and 95% of Ti$^{4+}$ ions are distributed over 4*h* sites. The Nd$^{3+}$ ions are distributed at 4*g* sites.

Figure 3 shows the magnetization curve taken at 0.1T. A sharp downturn in magnetization is observed at 11K. This is the marker for AFM ordering in both Nd$^{3+}$ and Cr$^{3+}$ ions [23]. On careful observation we observe a small bending at 18 K as well. This is the antiferromagnetic transition temperature corresponding to Cr$^{+3}$ ions, which has been confirmed by heat capacity measurements [10]. This feature is magnified in the inset 3(a). The inverse susceptibility ($\chi^{-1}$ = (T-θ)/C) is plotted in inset 3(b). This yields the Curie-Weiss constant C = 4.102 emu Oe$^{-1}$ mole$^{-1}$ K$^{-1}$ and the intercept θ = - 85.5K. The effective magnetic moment is found to be 5.728 $\mu_B$ which is close to theoretical value corresponding to one Nd$^{3+}$ (S = +3/2) and one Cr$^{3+}$ ions (S = +3/2) per formula unit. The M-H measurement at 2K (inset 3c) shows non-hysteretic behaviour (without saturation), that reflects the non- canting nature of the aligned spins below the ordering temperature.

To ascertain the magneto-electric coupling, next we discuss the measured dielectric constant as a function of temperature in the presence of magnetic field (Fig. 4). In the absence of magnetic field, slight bending in the curve around the transition temperature $T_N$ = 18 K is observed. As the magnetic field is applied, a clear peak appears at the transition temperature. The height of the peak at the transition temperature increases with magnetic field increment. We tried to find out the effect on the dielectric constant, when the sample is cooled in the presence of magnetic field but no additional features were observed. We also note that for higher fields, the peak maxima shifts to lower temperature. This is because the Neel transition temperature of antiferromagnets decrease with increasing field. Inset 4a amplifies the magneto-capacitance nature of $NdCrTiO_5$. On increasing magnetic field the capacitance increases. This change is most prominent around the AFM transition temperature and is substantially reduced at temperatures above and below $T_N$. We note that magneto-capacitance at 10K and 25K is negligible in comparison to that at 16K.

Next we address the question whether the induction of electric polarization necessarily requires the presence of magnetic field. The results from J. Hwang et al. could not conclude decisively on this because of the extremely low pyroelectric current magnitude in the absence of magnetic field. Fig. 5(a) shows the measured pyroelectric current as a function of temperature with ±200V/mm poling electric fields. It shows a consistently observed pyroelectric current of magnitude 0.2-0.3pA even in the absence of magnetic field. This current reverses its direction when the electric field direction is reversed. To establish this unambiguously, we have also measured pyroelectric current with temperature cycling. As shown in Fig. 5(b), we observe the direction change of the pyroelectric current during heating and cooling cycles [24, 25]. The temperature cycle is varied from ~15K to ~17K in a periodic manner and pyroelectric current changes sign following the profile of temperature variation. This pyroelectric current in the absence of magnetic field does not decay over time, which proves that it is associated with the movement of permanent dipole. This occurs due to preservation of polarized state below the transition temperature. When heat is supplied to polarized dipoles, their arrangement tends towards the low ordered state and this is recovered when specimen temperature is lowered. Such reversible phenomenon leads to generation of opposite current in intrinsic ferroelectric material. This emergent phenomena squarely places $NdCrTiO_5$ into the group of genuine multiferroic materials.

In Fig. 6 we plot the calculated polarization at a fixed magnetic field of 4T with varying poling electric fields. The corresponding pyroelectric currents are shown in inset 6a.

For a negative electric field, the pyroelectric current shows a peak in reverse direction with same magnitude. Inset (b) of Fig. 6 shows the polarizations at different magnetic fields. This confirms robust magneto-electric coupling in NdCrTiO$_5$. The magnetoelectric coefficient $\alpha_{ij}$ from the first order linear relation $P_i = \alpha_{ij}H_j$ is found to be of the order of ~ 0.43µC/m$^2$T. This is smaller than typical collinear multiferroic like Y$_2$CoMnO$_6$[9]. Though not shown here, we also investigated whether there is any spin capture or kinetic arrest of different magnetic phases [26] in which a threshold field is needed to break the spin symmetry. We did not observe such an effect.

**Conclusion:**

In conclusion, we find direct evidence for multiferroic behaviour in NdCrTiO$_5$. This is over and above large field dependence of electric polarization which is due to strong magneto-electric coupling. Our results suggest that below T$_N$ ~ 18K, the Cr$^{3+}$ ions become anti-ferromagnetically ordered that subsequently drives anti- ferromagnetism in Nd$^{3+}$ ions along *ab* plane through exchange coupling. Effectively no spin canting is observed in low temperature hysteresis measurements. A model based on collinearly ordered Cr$^{3+}$ions along c-axis giving rise to magneto-striction is proposed to explain multiferroicity in NdCrTiO$_5$. Such multiferroic compounds with large magnetic field dependence of electric polarization auger well for potential applications.

**Acknowledgement:**

The authors acknowledge Prof. A. K. Rastogi for discussion on magnetic properties of the sample. We thank AIRF JNU for XRD, SEM and magnetization measurements. JS thanks CSIR-Govt of India and GS thanks UGC-Govt of India for fellowships.

**Figure Captions:**

**Fig. 1.** (Color online) Room temperature XRD pattern of NdCrTiO$_5$ polycrystalline powder sample with Reitveld refinement. The observed data (red), calculated line (black) and difference (blue) between the two are shown along with the Bragg positions (green). Orthorhombic crystal structure in space group *Pbam* is confirmed.

**Fig. 2.** (Color online) (a) Schematic of unit cell orthorhombic crystal structure of NdCrTiO$_5$ as seen along the c axis. Nd$^{+3}$, Cr$^{+3}$ and Ti$^{+4}$ ions are shown in dark yellow, blue and cyan respectively. (b) The unit cell as seen from the basal plane.

**Fig. 3.** Magnetization of NdCrTiO$_5$ measured in presence of 0.1T magnetic field between temperatures 5K to 300K. The inset (a) shows magnified low temperature data. Inset (b) shows the inverse susceptibility ($\chi^{-1}$) as a function of temperature with the Curie-Weiss fitting. Inset (c) shows the M-H hysteresis curve at T = 2K.

**Fig. 4** (Color online) Dielectric constant is plotted as a function of temperature. A large increase is seen with increasing magnetic field. Inset shows the capacitance versus magnetic field at different temperatures 10K, 16K and 25K. Large magneto-capacitance is observed just below the Neel Temperature.

**Fig. 5.** (Color online) (a) Pyroelectric current as a function of temperature at H = 0T. The direction of current reversal is also confirmed for negative poling field (b) Pyroelectric current as a function of time with cyclic change of temperature. Such variation can only be observed because of reversible ferroelectric domain condensation.

**Fig. 6.** (color online) Electric polarization is plotted as a function of temperature at ± 200 V poling field. A constant magnetic field of 4 Tesla is applied while warming. Inset (a) Pyroelectric current as a function of temperature at ± 200 V poling field and H= 4T. Inset (b) Polarization as a function of temperature at various magnetic fields.

**Table 1.** Structure parameters of NdCrTiO$_5$

| Atom | X | Y | Z |
|---|---|---|---|
| Nd | 0.14091 | 0.17045 | 0 |
| Cr | 0.11730 | -0.14443 | 1/2 |
| Ti | 0 | ½ | 0.25319 |
| O1 | 0.10715 | -0.28281 | 0.26618 |
| O2 | 0.16262 | 0.44428 | 0 |
| O3 | 0.15288 | 0.43262 | 1/2 |
| O4 | 0 | 0 | 0.30046 |

| Bond | Length(Å) |
|---|---|
| Nd-O1 | 2.486 |
| Nd-O2 | 2.375 |
| Nd-O2 | 2.457 |
| Ti-O1 | 1.811 |
| Ti-O3 | 1.836 |
| Ti-O4 | 1.921 |

| Bond | Angle (Degree) |
|---|---|
| Nd-O1-Ti | 122.696 |
| Nd-O1-Cr | 97.554 |
| Nd-O2-Cr | 101.557 |
| Ti-O3-Cr | 131.962 |
| Nd-O4-Ti | 127.018 |

**Figure1.**

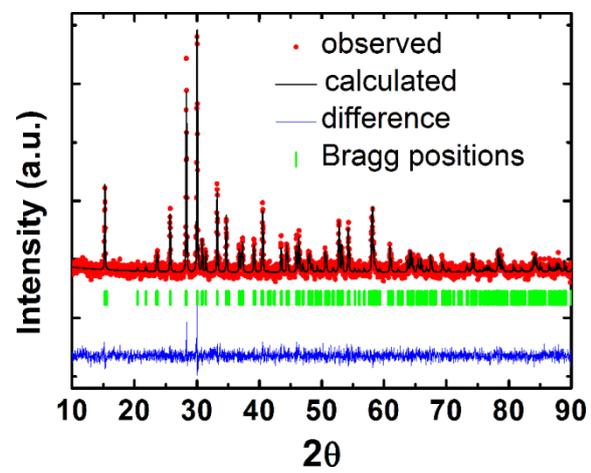

**Figure2.**

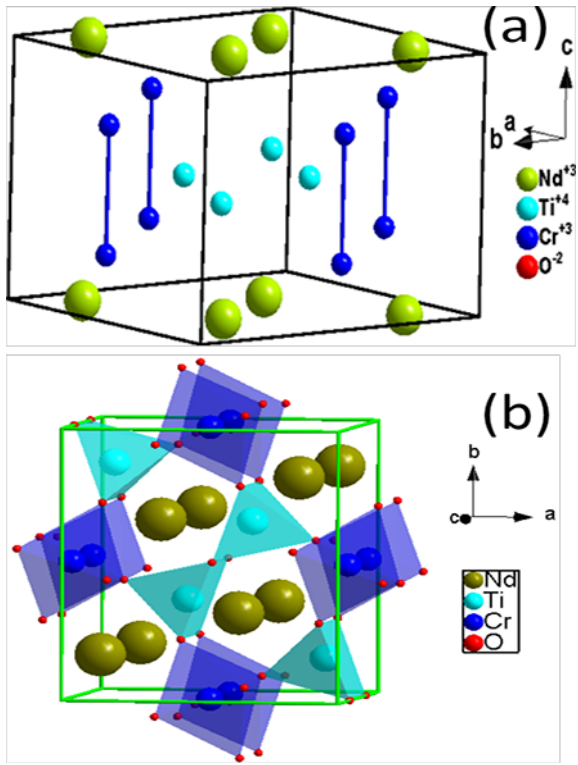

**Figure 3.**

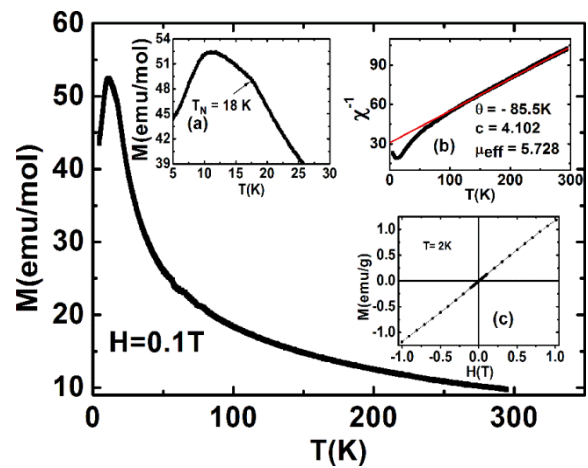

**Figure4.**

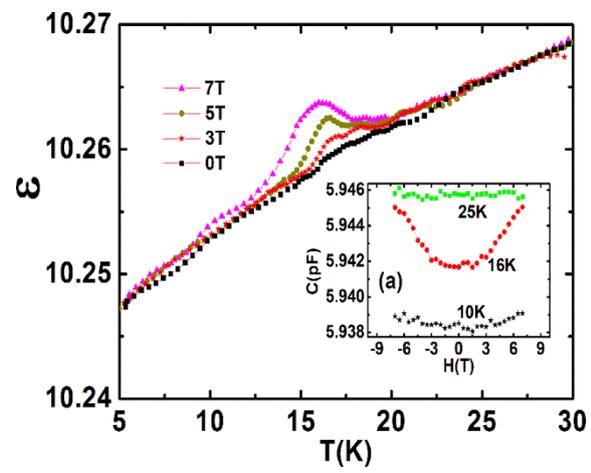

**Figure5.**

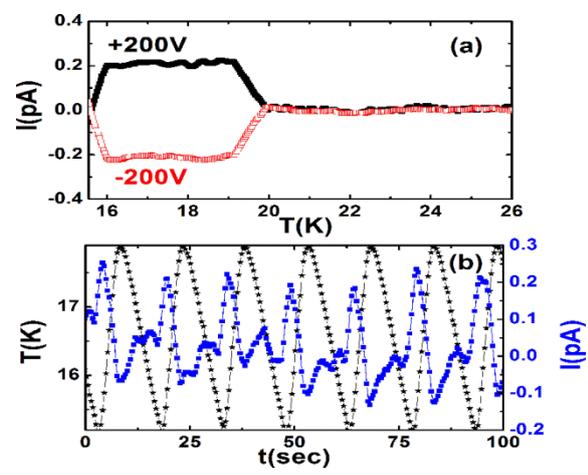

**Figure6.**

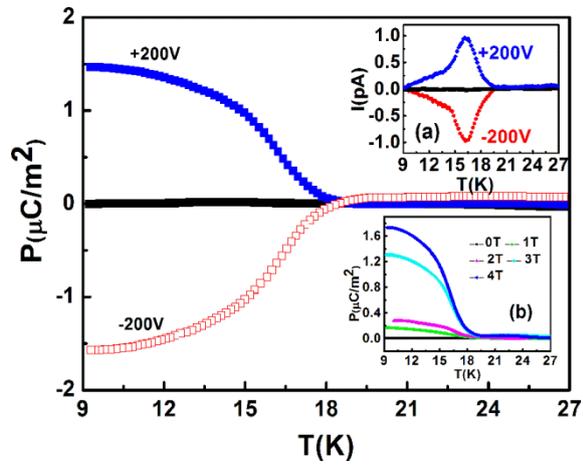